\documentclass{article}
\usepackage{arxiv}
\usepackage[utf8]{inputenc} 
\usepackage[T1]{fontenc}    
\usepackage{hyperref}       
\usepackage{url}            
\usepackage{booktabs}       
\usepackage{amsfonts}       
\usepackage{nicefrac}       
\usepackage{microtype}      
\usepackage[ruled,vlined]{algorithm2e}
\usepackage{amssymb,amsbsy, amsmath}
\usepackage[dvipsnames,usenames]{color}
\usepackage{graphicx}
\usepackage{rotating}
\usepackage{dcolumn}
\usepackage[normalem]{ulem} 
\usepackage{xr}
\usepackage{nameref}
\usepackage[square, comma, numbers, sort&compress]{natbib}

\usepackage{zref-xr}
\zxrsetup{toltxlabel}

\usepackage{setspace}

\newcommand*\diff{\mathop{}\!\mathrm{d}}

\title{Simulation-Based Inference with Approximately Correct Parameters via Maximum Entropy}

\author{%
  Rainier Barrett\\
  Department of Chemical Engineering\\
  University of Rochester\\
  Rochester, NY 14627\\
  \And
  Mehrad Ansari \\
  Department of Chemical Engineering\\
  University of Rochester\\
  Rochester, NY 14627\\
  \And
  Gourab Ghoshal \\
  Department of Physics and Astronomy\\
  University of Rochester\\
  Rochester, NY 14627\\
  \And
  Andrew White\thanks{andrew.white@rochester.edu} \\
  Department of Chemical Engineering\\
  University of Rochester\\
  Rochester, NY 14627
}

\begin{document}

\maketitle
\begin{abstract}
\singlespacing
Inferring the input parameters of simulators from observations is a crucial challenge with applications from epidemiology to molecular dynamics. Here we show a simple approach in the regime of sparse data and approximately correct models, which is common when trying to use an existing model to infer latent variables with observed data. This approach is based on the principle of maximum entropy (MaxEnt) and provably makes the smallest change in the latent joint distribution to fit new data. This method requires no likelihood or model derivatives and its fit is insensitive to prior strength, removing the need to balance observed data fit with prior belief. The method requires the ansatz that data is fit in expectation, which is true in some settings and may be reasonable in all with few data points. The method is based on sample reweighting, so its asymptotic run time is independent of prior distribution dimension. We demonstrate this MaxEnt approach and compare with other likelihood-free inference methods across three systems; a point particle moving in a gravitational field, a compartmental model of epidemic spread and finally molecular dynamics simulation of a protein.
\end{abstract}

\section*{Introduction}
Simulation-based inference (SBI) is a class of methods that infer the input parameters and unobservable latent variables in a simulator from observational data. SBI is different than traditional statistical inference or machine learning because simulators are typically not differentiable and their likelihoods are intractable. There have been great strides in methods for SBI and a recent review may be found in~\cite{Cranmer2020}. Most SBI methods are concerned with finding a few simulator parameters from a rich set of observations~\cite{rubin1984bayesianly, beaumont2002approximate, diggle1984monte}. Here, we consider updating a simulator with many trusted parameters to match a sparse set of observations. The ancestor for this line of research is in molecular dynamics simulations of proteins. These simulations require thousands of parameters and the observed data (macroscopic experimental values) is often on the order of 10 to 100 data points (e.g. Reißer et al.\cite{BussiRNA}). An approach that has emerged in molecular dynamics simulations is maximum entropy (MaxEnt) biasing~\cite{Sormanni2017,bonomi2017principles, Olsson2013,Amirkulova2019}. MaxEnt biasing minimially modifies the simulator to match observations. The premise of MaxEnt is that the original model is approximately correct and observations should be matched in expectation, which is not the usual approach in SBI. These two assumptions lead to a unique bias~\cite{Pitera2012} to the simulator that is independent of the parameters and can be implemented as a simple reweighting procedure. The MaxEnt method's run-time scales only with sample number, rather than the number of model parameters which is atypical of most SBI methods because they require joint sampling.

Our MaxEnt method reweights a black-box simulator to agree with observed data in a provably minimal way. The reweighted simulator can then be used to infer either better input parameters or other simulation outputs. The two conditions are that (i) the simulator is accurate enough that the observed data could have been derived from an average of runs of the simulator; and (ii) predicted values for the observed data can be computed from the outcome of the simulator. The MaxEnt method results in an ensemble of outcomes from the simulator whose means agree with data and provide a regressed agreement to observed data while being as close to the original simulator outcomes as possible. The method is efficient, provides uncertainty estimates, and can account for unknown systematic errors.

This paper focuses on a setting where distribution moments (e.g., population average) are the data for fitting a posterior. This is a common setting of MaxEnt and it has a number of advantageous properties. Finding the MaxEnt posterior is equivalent to maximizing the likelihood function under a distribution family (exponential in this work)\cite{berger1996maximum}. The MaxEnt posterior is the closest to the prior distribution (under KL divergence) under the constraint of fitting the population averages\cite{Roux2013statistical}. The MaxEnt posterior exactly fits the distribution moments under mild assumptions\cite{Pitera2012}. \footnote{Bayesian inference for fitting distribution moments requires specifying an error distribution that requires additional system insight and its strength relative to the prior belief affects the agreement with the distribution moment data. See system 1.} Examples of MaxEnt in this setting can be seen in statistical mechanics as described above, biology\cite{de2018introduction}, natural language processing\cite{berger1996maximum}, and ecology\cite{banavar2010applications}. Any application of maximum likelihood on distribution moments can be recast as MaxEnt. Our contribution to this setting is to summarize the general theory and provide an efficient and simple implementation that is system independent.

The second setting of MaxEnt is to make an ansatz that an observation can be substituted as a distribution moment. For example, consider observing a particle trajectory and we would like to make inferences about where the particle will go next. Our observations are specific pairs of time and position that describe the trajectory. If we have a good model for how the particle behaves, MaxEnt will minimally change the model to agree with the particle trajectory \textit{on average}. The MaxEnt posterior will agree in expectation exactly with the observed trajectory. Thus, from a practical point of view MaxEnt provides an accurate description of the data and a probability distribution for the posterior. The inferred continuation of the trajectory will come from the expectation of that posterior. An alternative would be treating the observation as exact and regressing the prior model, which would not give a posterior but instead a mode (most likely trajectory). Yet this gives no uncertainties with the predictions and can lead far away from the prior model, leading to issues like overfitting and covariate shift. Bayesian inference could be used to fit the particle trajectory by supposing a measurement error distribution. Yet this creates an over-constrained problem where a weak error distribution reduces the agreement with the observed trajectory and a strong error distribution reduces the agreement of the prior model. In essence, by "relaxing" the observation to be an average we enable agreement with the observation exactly, maximize the agreement with the prior, and do not require potentially ad-hoc construction of error distributions\footnote{Our maximum entropy formulation does allow uncertainty on distribution moments if desired.}. This can be justified through the principle of maximum parsimony: the MaxEnt formulation requires the fewest input parameters. If multiple observations are gathered, then the Bayesian inference setting is more appropriate because the distribution moment ansatz would exclude information about variance in multiple observations. Another potential application area could be in few-shot regression with Bayesian network models\cite{wilson2020bayesian}, where only a few examples are available in a new task. MaxEnt provides a way to fit a previously trained Bayesian network to those few examples balancing agreement with them exactly and while minimizing the effect on the trained model.

The MaxEnt method presented here has a run time scaling that is independent of the number of model/prior parameters; it acts entirely on samples. This also means that intractable or infinite dimensional priors (such as sampling both models/priors) can be treated with MaxEnt. This can be a large advantage over other approximate inference methods like approximate Bayesian computing and likelihood free inference.

The MaxEnt approach in simulation can be traced to Jayne's early work on deriving statistical physics from MaxEnt~\cite{Jaynes1957MaxEnt}. It was shown, for example, that the Boltzmann distribution could be derived by simply adding a restraint on average energy that must be satisfied in expectation, analogous to matching an observation. A similar method of incorporating observations in expectation returned 50 years later in determining how to match protein molecular dynamics simulations to observations~\cite{islam2013structural}. This method was then recast as an approximation to MaxEnt~\cite{Roux2013statistical}. Matching observations in molecular dynamics with MaxEnt was also shown in Pitera and Chodera~\cite{Pitera2012}. This was followed by rapid progress to create practical methods for use in simulations~\cite{Vendruscolo2013molecular, Boomsma2014combining, white2014efficient, Amirkulova2019}. The MaxEnt method based on reweighting has been presented in the context of molecular dynamics simulations in many forms over the years~\cite{BussiRNA, beauchamp2014bayesian, rozycki2011saxs, leung2016rigorous, choy2001calculation, bernado2007structural, berlin2013recovering, bertini2010conformational, pelikan2009structure, shaw2010atomic}.  MaxEnt-based methods have a long history of use in the molecular dynamics community across various types of systems, and this approach is still widely used for biasing applications in modern molecular simulations, demonstrating ongoing interest and engagement in the community\cite{Bottaro2020BayesMaxEnt,Bradshaw2020DeuteriumMaxEnt,Lou2018HistogramMaxEnt}.  A review by Bonomi et al. provides broader context for the use of maximum entropy and other similar methods in the molecular simulation community~\cite{bonomi2017principles}. The review by Cesari, Reißer, and Bussi\cite{Cesari2018MaxEntHowTo} provides an overview of the mathematics of MaxEnt, its connections to Bayesian inference and maximum likelihood, and some discussion of the potential hurdles involved. Also, for a comparative study weighing the benefits of MaxEnt and restraint-based methods, see Rangan et al. 2018\cite{Rangan2018RestraintvsReweight}.

Our contribution here is deriving a general MaxEnt framework that is applicable to arbitrary simulators, demonstrating its application to areas outside of molecular dynamics, and showing one method of improving the support (sampling) of the posterior, which is important when the simulator is far from the observations. In the remainder of this work, we develop the theory, discuss sampling issues, and compare the MaxEnt method to approximate Bayesian computing (ABC)~\cite{beaumont2002approximate,Blum2010SIRABC,Toni2009SIRABC,Kypraios2017SIRABC}, sequential neural likelihood (SNL)~\cite{papamakarios2019SNLE}, and direct Bayesian inference when the likelihood is tractable. Additional background on these methods used for comparison can be found in the supporting information.

\section*{Results}
\label{results}

\subsection*{Theory}
Given a simulator $f(\vec{\theta})$ with a set of parameters $\vec{\theta}$, we have a prior distribution of parameters $P(\vec{\theta})$. For example, the function $f(\vec{\theta})$ could be propagating a system of ODEs for some set number of timesteps or a molecular dynamics simulation with intrinsic noise.

Suppose we have some set of $N$ observations, $\{\bar{g}\}_k$, $k\in[1,\ldots,N]$, which we would like to match with our model. Assume the measurement of each $\bar{g}_k$ has some uncertainty $\epsilon_k$, where $\epsilon_k$ is a random variable distributed according to some prior distribution about uncertainty, $P_0(\epsilon_k)$. We would like to constrain our model such that
\begin{equation}
    \int\diff\vec{\theta}\diff\vec{\epsilon} P'(\vec{
    \theta}) P_0(\epsilon_k) \left( g_k[f(\vec{\theta})] + \epsilon_k \right) = E[g_k + \epsilon_k] = \bar{g}_k~ \forall k
\end{equation}
This means that we want the average over the distribution of our updated models ($P'(\vec{\theta})$) to match the observations data, with an allowable average disagreement based on $\{\epsilon_k\}$. This is an unusual constraint and is weaker than most simulation inference methods. It reflects the strong belief in our prior model in this setting. Note that inclusion of the $P_0(\epsilon_k)$ and $\epsilon_k$ terms is optional: it is not necessary to allow disagreement on average with data, unlike in a Bayesian framework. This would be equivalent to setting the error distribution to a Dirac delta about 0: $P_0(\epsilon_k) = \delta(\epsilon_k = 0)$. Another difference is that this distribution of uncertainty is about bias. It accounts for systemic deviation in average agreement and does not describe the underlying variance of the observational data. This approach is analogous to Bayesian model averaging~\cite{Gordon2020BayesModelAverage}, in that it is an average over many model parameter settings, reweighted by the posterior likelihood.

The maximum entropy modification to the prior distribution $P(\vec{\theta})$ to satisfy the N constraints is given by~\cite{Roux2013statistical,Pitera2012,Amirkulova2019,cesari2016MaxEntUncertainty}:

\begin{eqnarray}
  P'(\vec{\theta}, \vec{\epsilon}) &= \frac{1}{Z'}P(\vec{\theta})\prod_k^Ne^{-\lambda_kg_k[f(\vec{\theta})]}e^{-\lambda_k\epsilon_k}P_0(\epsilon_k), \\
  Z' &= \int\diff \vec{\theta} \diff \vec{\epsilon} P(\vec{\theta}) P_0(\epsilon) e^{-\sum_k\lambda_k (g[f(\vec{\theta})] + \epsilon_k)},
\end{eqnarray}
where $Z'$ is a normalization constant and $\lambda_k$ are chosen such that $E[g_k + \epsilon_k] = \bar{g}_k$. The dependence on  ${\vec{\epsilon_k}}$ can be removed by computing the marginal,\begin{equation}
        \label{eq:maxent}
        P'(\vec{\theta}) = \int \diff \vec{\epsilon}P'(\vec{\theta}, \vec{\epsilon}) = \frac{1}{Z'}P(\vec{\theta})\prod_k^Ne^{-\lambda_kg_k[f(\vec{\theta})]}\int \diff \epsilon_k e^{-\lambda_k\epsilon_k}P_0(\epsilon_k).\end{equation}
The problem is reduced to finding $\lambda_k$ such that the constraint is satisfied. Again, we must remove $\vec{\epsilon_k}$ from $E[g_k] + E[\epsilon_k] = \bar{g}_k$, where
$E[\epsilon_k]$ is:
\begin{equation}
\label{eq:expected_epsil}
    E[\epsilon_k] = \frac{\int \diff \epsilon_k e^{-\lambda_k\epsilon_k}P_0(\epsilon_k)\epsilon_k}{\int \diff \epsilon_k  e^{-\lambda_k\epsilon_k}P_0(\epsilon_k)},
\end{equation}
and is understood to still be a function of $\lambda_k$. If we define $\xi_k(\lambda_k) = E[\epsilon_k]$  the constraint equation can be rewritten as $E[g_k] + \xi_k(\lambda_k) = \bar{g}_k$. If the prior is an exponential family, the $\lambda_k$s will exist and be unique under some mild assumptions about support of the prior and covariance of observables (i.e., cannot have perfectly correlated observables with incompatible observations)\cite{Pitera2012,Roux2013statistical}.

\subsection*{Computing weighted properties and sampling efficiency}
In Algorithm \ref{alg:maxentweights}, we show the procedure for sampling from the MaxEnt distribution defined in Equation \ref{eq:maxent} via importance sampling~\cite{Tokdar2010ImportanceReview}. Here, $P(\vec{\theta})$ is the prior distribution over simulation parameters $\vec{\theta}$, $f$ is the simulator, $M$ is the number of samples from the prior to take, $N$ is the number of constraints, $P_0(\epsilon_k)$ is the error distribution, $g_k$ is the $k$th observation, and $\bar{g}_k$ is the target observable value to which we would like to constrain our simulator. $\eta$ is the learning rate. The output of this algorithm, $\{w_i\}$, are the weights of trajectories $\{f(\theta_i)\}$, and any desired property $g$ can be computed as $\sum_i g[f(\vec{\theta}_i)]w_i/\sum_iw_i$.

\begin{algorithm}[t!]
    \label{alg:maxentweights}
    \SetAlgoLined
    Input $P(\vec{\theta})$, $f(\vec{\theta})$, $M$, $N$ of $P_0(\epsilon_k)$, $g_k$, $\bar{g}_k$, $\eta$\\
    Initialize $\lambda_k = 0$ $\forall k$\\
    \For{$i\gets1$ \KwTo $M$}{
        Sample $\vec{\theta}_i \sim P(\vec{\theta})$\\
        \For{$k \gets 1$ \KwTo $N$}{
        Evaluate $g_k[f(\vec{\theta}_i)]$ \\
        }
    }
    \While{$\sum_i w_ig_k[f(\vec{\theta_i})] / \sum_i w_i + \xi_k(\lambda_k) \neq \bar{g}_k$ for any $k$}{

        \For{$i\gets1$ \KwTo $M$}{
            $w_i \gets \prod_k e^{-\lambda_k g_k[f(\vec{\theta_i})]}\int \diff \epsilon_k e^{-\lambda_k\epsilon_k}P_0(\epsilon_k)$\\
            \For{$k\gets1$ \KwTo $N$}{
            $\lambda_k \gets \lambda_k - \eta \frac{\partial}{\partial \lambda_k}\left( \sum_l \left( \bar{g}_l - \left[g[f(\vec{\theta}_i)] w_i / \sum_j w_j + \xi_l(\lambda_l)\right]\right)^2\right)$
            }

        }

    }
    \Return $\vec{w}$
    \caption{MaxEnt Weights with Uncertain Observations}

\end{algorithm}

The challenge of using MaxEnt is sampling from $P'(\vec{\theta})$. Our assumption thus far is that our prior $P(\vec{\theta})$ is approximately correct, so that samples  from $P(\vec{\theta})$ should be similar to $P'(\vec{\theta})$. In this ideal case, the algorithm is simply a matter of reweighting. One samples $\vec{\theta}_i$, computes $f(\vec{\theta}_i)$, compute weights proportional to $w_i[P'] = \prod_k^Ne^{-\lambda_kg_k[f(\vec{\theta})]}$ consistent with the experimental data (Algorithm \ref{alg:maxentweights}), and then any other property is reweighted with the same weights. In the non-ideal case (if for instance sampling is expensive, the space is high-dimensional, or the model is far from correct), there can be insufficient support to agree with the constraints. To treat insufficient support, we take a simple approach and use gradient descent to modify the sampling distribution parameters $\vec{\theta}^j$ to minimize the cross-entropy with $P'(\vec{\theta})$:

\begin{equation}
\label{eq:sampler}
    \vec{\theta}^{j+1} = \vec{\theta}^j - \eta \nabla_{\vec{\theta}^j} \sum_iw_i[P']\ln P(\vec{\theta}_i),
\end{equation}

where $w_i[P']$ depends on $\vec{\theta}^j$ via the expectation function. We remove the effect of the sampling distribution from the posterior via reweighting by $P(\vec{\theta}) / P(\vec{\theta}^j)$ We refer to this approach as \textit{variational}.

\subsection*{Trivial Simulation with Gaussian Noise}
\begin{figure}[t!]
    \centering
    \includegraphics[width=\textwidth]{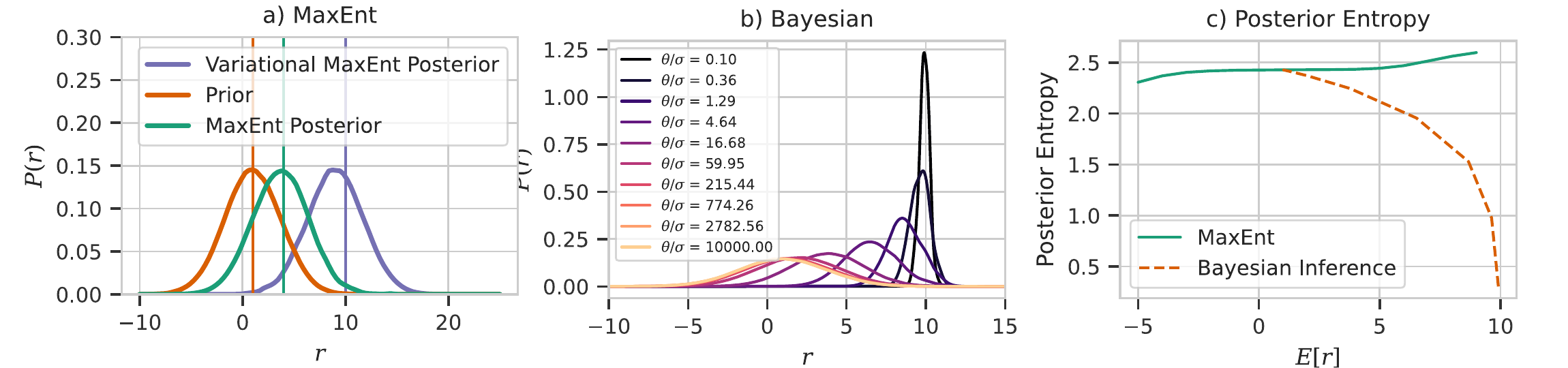}
    \caption{{\bf Comparison of Bayesian inference and MaxEnt reweighting}. Panel a) shows the simulator prior distribution in orange and the two versions of MaxEnt posterior with observations of 5 and 10. Panel b) shows the interplay between strength of prior and assigned uncertainty to the observation at 10 for Bayesian inference. The value is arbitrary and chosen to illustrate how Bayesian inference strongly alters the shape of the posterior compared to the prior, whereas MaxEnt preserves the shape well. Note the scale change between panels a) and b). Panel c) further illustrates this by comparing the posterior entropy of the two as a function of the observation location.\label{fig:compare}}
\end{figure}

We first consider a toy simulator $f$ that outputs a scalar $r$. We have a prior belief about the value of the constant as a normal distribution $\mathcal{N}(\hat{r}, \theta)$. This example serves to compare the MaxEnt approach with Bayesian inference. The observed data is a single point ($\bar{r}$) and we treat it as an average constraint in the MaxEnt. That is, we have a single observation and we constrain our simulator to on average match this observation. Figure~\ref{fig:compare} panel a) shows how the MaxEnt posterior changes with different observations ($\bar{r} = 5$ or $\bar{r} = 10$). The $r = 10$ observation requires the variational sampling (Equation~\ref{eq:sampler}) because the observed value is outside the sampled support of the prior. The expected value of $E[r]$ of the posterior always matches the observation and the moments of the posterior are identical to the prior, except the 1$^\textrm{st}$ moment (the mean). Although Figure~\ref{fig:compare}a is calculated with Algorithm~\ref{alg:maxentweights}, the analytic equation for the posterior is simply $\mathcal{N}(\bar{r}, \theta)$\cite{Pitera2012}.

With Bayesian inference, we must assume some noise model of our simulator so that we can compute the probability of the single observation arising from the simulator, namely $P(\textrm{data} | \textrm{model})$\cite{hummer2015bayesian}.
We take this to be $\mathcal{N}(\hat{r}, \sigma)$. The Bayesian posterior balances this evidence with the prior distribution:
\begin{equation}
    P_B\left(r\right|\left.\bar{r}\right) = \frac{1}{Z} e^{-\frac{\left(r - \bar{r}\right)^2}{2\sigma}} e^{-\frac{\left(r - \hat{r}\right)^2}{2\theta}}
\end{equation}


The expected value of $\hat{r}$ will not match the observed value except in the limit of $\sigma / \theta$ reaching 0. $E[\hat{r}]$ will be between the observed value and the prior belief expectation.  Figure~\ref{fig:compare}b compares the Bayesian inference and MaxEnt posteriors.
Panel a) shows how the MaxEnt method leaves the variance of $\hat{r}$ unchanged as we consider different observed values.
Panel b) shows the use of Bayesian inference to match the observation at $r = 10$. It requires extreme ratios between prior belief and experimental uncertainty to match the observation at 10. This is not necessarily a disadvantage, we simply are showing that observations are matched in expectation with MaxEnt and not with Bayesian inference. Panel c) shows how the MaxEnt method keeps the posterior entropy maximized regardless of the observation value (x-axis), as expected for a maximum entropy method. Bayesian inference shows a more peaked distribution when the observed value is far away from the prior, giving less entropy. That is, to agree with the observation we must necessarily increase the strength of evidence, which peaks the posterior.

\subsection*{Example System 1: Point Particle Gravitation Simulation}
\begin{figure}[t!]
    \centering
    \includegraphics[width=0.95\textwidth]{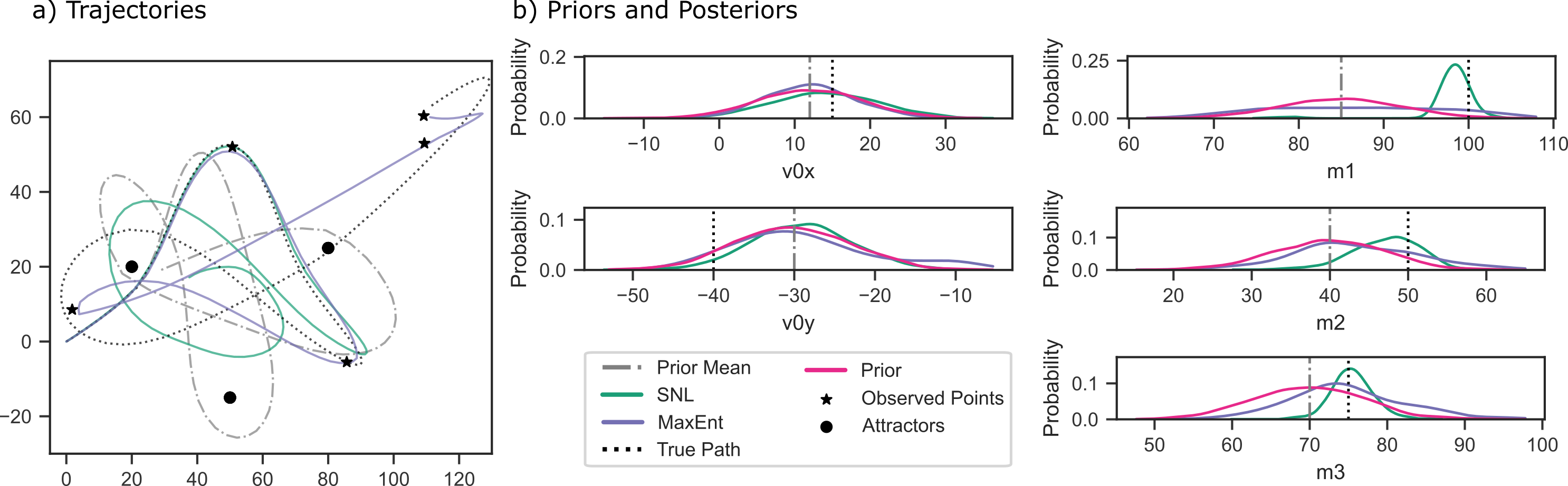}
    \caption{
    \textbf{Comparison of SNL and MaxEnt methods on a gravitational field simulation of a particle moving through a fixed field with three attractors.} All units are SI values (m, m/s, kg). \textbf{a)}: weighted mean paths generated by SBI with SNL (green) and MaxEnt (purple), alongside the path generated by the mean of the prior distribution (dash-dotted grey), and the true path used to generate observations (dashed black). Target points appear as black stars, and the attractors are black circles. \textbf{b)}: Kernel density estimate of the posterior distribution of parameters after fitting, alongside their respective priors.}
    \label{fig:gravitation}
\end{figure}

For a our first example system, Figure \ref{fig:gravitation} shows a comparison of SNL and MaxEnt reweighting on a unit mass particle in a gravitational field of three attractors. The simulator here is a point particle following Newtonian gravitational mechanics. The goal here is to modify the simulation trajectories to align with a small set of observations. An example task might be fitting the trajectory of a comet to a small number of observations separated by years.

The parameters for this simulation were $m_1$, $m_2$, $m_3$, $v_{0x}$ and $v_{0y}$, the masses of the three attractors, and the initial velocity of the particle, respectively. The positions of the attractors and the initial position of the particle were all fixed. We treat these parameters as unknown, and the prior belief for them follows a  normal distribution, shown in Figure~\ref{fig:gravitation}. Repeatedly sampling from this prior and running the simulator results in a distribution of trajectories, whose means are shown in Figure \ref{fig:gravitation}a). MaxEnt reweights this ensemble of trajectories to agree with five observed positions along the trajectory. (The mean path does not exactly pass through the observations because some zero-mean normally-distributed noise with standard deviation of 3 was added to each observation.) The average posterior trajectory indeed agrees with all observations. The prior and posterior for the parameters are shown in Figure \ref{fig:gravitation}b.

The observed points were synthesized by choosing a set of true parameter values and imposing zero-mean normally-distributed noise with standard deviation 3 on every 20th timestep on the 100-step simulation. Thus, one way to evaluate the MaxEnt performance is to see if the posterior means are close to these true values. We can see that the MaxEnt posteriors are closer to these values than the prior, but still largely in agreement with the prior. It fits the observations while staying as close to the prior as possible, because that maximizes entropy. In contrast, SNL results in a much narrower posterior around the true values, while diverging from the prior, because that maximizes likelihood.

We computed the cross-entropy of the prior and posterior produced by MaxEnt and SNL. These values were 5.09 for SBI with SNL, and  3.43 for MaxEnt reweighting. This demonstrates how MaxEnt minimally alters the prior distribution while still matching observations in expectation -- the average path followed by the MaxEnt particle matches all target points, while matching the posterior to the prior's shape more closely than SNL.

 This example illustrates two key points. Firstly, it shows that MaxEnt is robust to chaotic systems, as it is still able to match observations on average, with minimal change to the prior. However, it is also an example of when another SBI method may be preferable, depending on the goal. The goal of MaxEnt is, by construction, to alter the prior as little as possible while agreeing with observations on average. In cases where the true underlying parameters governing a model are in a low-density region of the prior, the posterior resultant from MaxEnt will therefore assign relatively low probability to these parameter values as well. Thus, in the sense of estimating likelihood when the prior is not close to the true values, other methods like SNL can be preferable to MaxEnt. We can see that while SNL makes a better estimate of the true parameters used to generate those observations, it does not reproduce a path that aligns with the observations. This presents a choice to the simulator. If the goal is to alter a model to agree with observations, MaxEnt is preferred, especially if the prior is strongly trusted. If the goal is accurate likelihood estimation, methods like SNL are preferred, especially if the prior is not strongly trusted.

\subsection*{Example System 2: Epidemiological modeling}
\begin{figure}[h]
    \centering
    \includegraphics[width=4 in]{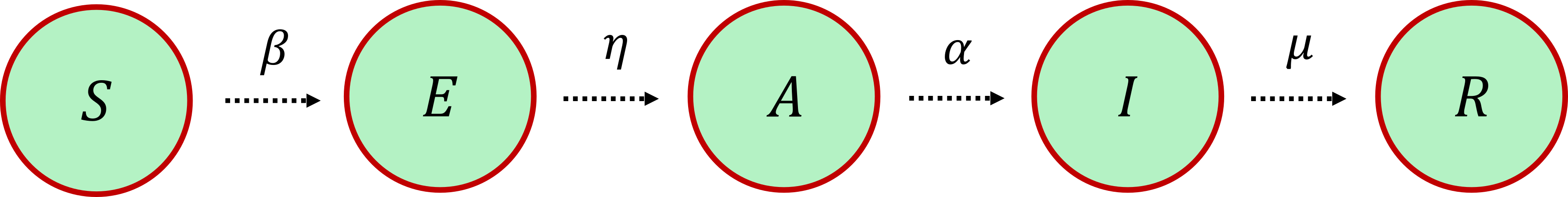}
    \caption{\textbf{SEAIR model.} Populations in each patch can be in any one of Susceptible ({\bf S}), Exposed ({\bf E}), Asymptomatic ({\bf A}), Infected ({\bf I}) and Resolved ({\bf R}). Susceptible individuals can get exposed to the disease by having contacts with the asymptomatic or the infected at infectivity rate $\beta$. Once exposed, they become asymptomatic and infected at rates $\eta$ and $\alpha$. The infected finally recovers or dies at rate $\mu$ and becomes resolved.   \label{fig:SEAIR}}
\end{figure}

In our third example, we apply our framework to modeling the spread of a pathogen in vulnerable populations. We consider an SEAIR compartmental model of epidemic spread (Figure \ref{fig:SEAIR}) on metapopulations connected via a spatial network of patches. Each patch corresponds to a location such as a zipcode in a city, or a county, and connections between patches correspond to mobility flows of residents encoded in a $M \times M$ mobility matrix for $M$ patches, where $M_{ij}$ is the number of people moving from patch $i$ to patch $j$ in one time increment. Contacts within patches occur in a fully-mixed mean field manner where individuals can be in any one of five states of infection: Susceptible ({\bf S}), Exposed ({\bf E}), Asymptomatic ({\bf A}), Infected (I), and Resolved ({\bf R}). The choice for this particular combination of compartments was inspired by its relevance in modeling the evolution of the current SARS-CoV-2 pandemic~~\cite{zhou2020pneumonia,wu2020new}. Each individual patch is represented with fractions of \textbf{S, E, A, I, R}, rather than the count of individuals within each compartment.

\begin{figure}[b!]
    \centering
    \includegraphics[width=1\textwidth]{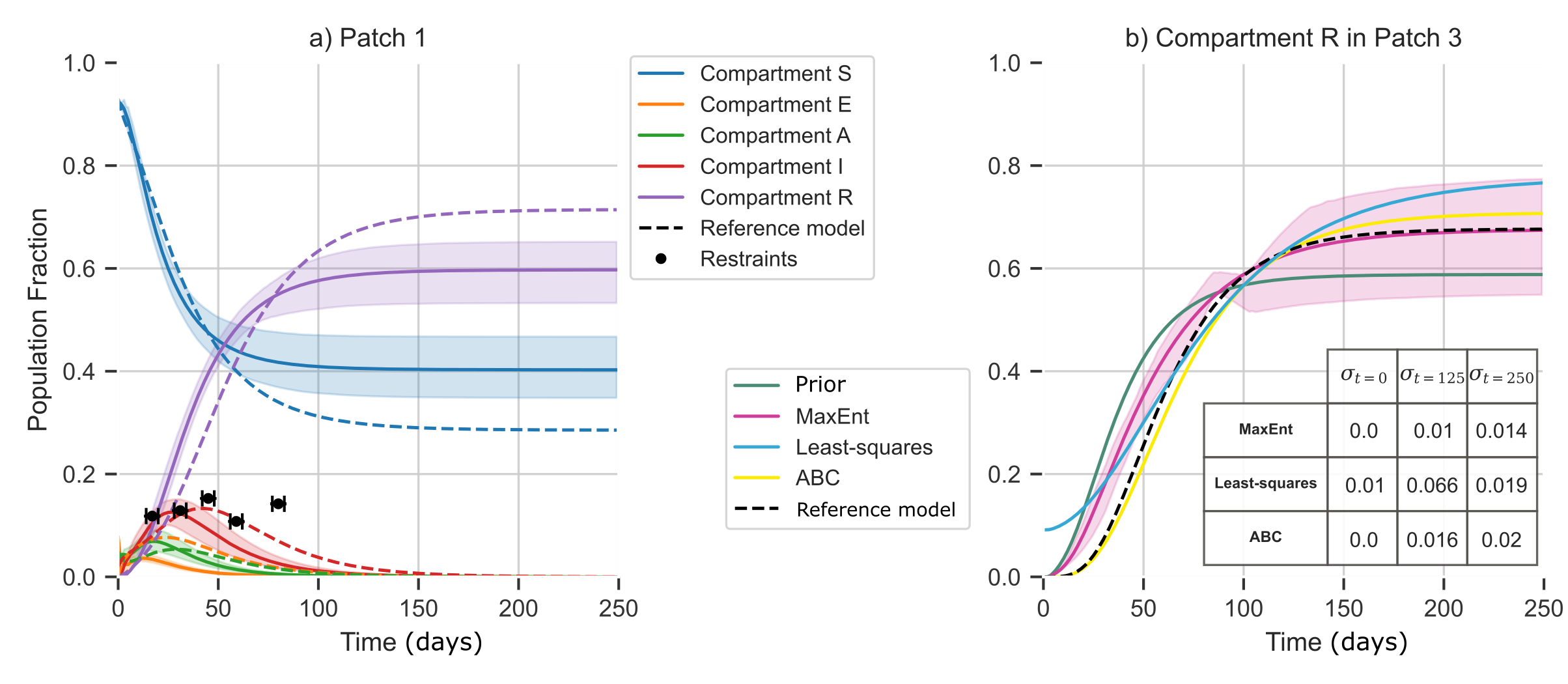}
    \caption{{\bf Maximum entropy reweighting of disease trajectory in a meta-population SEAIR model}. \textbf{a)}: Prior-generated trajectory in one of the spatial patches for compartments S (blue), E (orange), A (green), I (red) and R (purple) are shown with solid lines and the reference trajectory is in dashed lines. The colored area represents the one-third higher and lower quantiles than average. Observations shown as restraints (black circles) are selected randomly from compartment I with 5 percent additive noise and Laplace prior of 0.01. \textbf{b)}: Comparing the performance of MaxEnt (pink), Least-squares (blue), ABC (yellow) in fitting to reference model (black dashed line) in patch 3, based on observations in patch 1. Table inset shows standard deviations from 5-fold cross validation of the observations at three different times. Shading on MaxEnt is from $\pm$67\%  weighted quantiles.}
    \label{fig:Figure_4}
\end{figure}

We first create a ``reference'' trajectory that represents the true disease model. From this reference trajectory, we extract observations which are used as the input to the MaxEnt methods, by extracting values at specific timepoints in the reference trajectory. A challenge in modeling the spread of epidemics is associated with reporting of the empirical number of confirmed cases (compartment {\bf I}), which is typically very noisy~\cite{Lipsitch2020COVIDLetter}. To simulate this uncertainty, we add random additive noise to the observations from the reference trajectory (see Methods for details). This reference trajectory is represented as dashed lines in Figure \ref{fig:Figure_4}a). We choose 5 uniformly random data points within the first half of the trajectory of the compartment {\bf I} in patch 1 as observations (represented as black dots). The performance of the model is evaluated by comparing the predicted trajectory and the reference in a different patch (3). In Figure \ref{fig:Figure_4}b) we compare the performance of MaxEnt, a least-squares fit, and ABC in fitting the prior to the observations.
Compared to MaxEnt, the result from the least squares method was a poor fit with high variance, as it over-fits to observation noise. This was shown by doing a 5-fold leave-one-out cross-validation of the observations and evaluating the standard deviation at times $t = 0, 125$ and $250$ for each method (inset in Figure \ref{fig:Figure_4}.b). Out of all methods evaluated, ABC had the least variance, but was computationally more expensive to run, whereas MaxEnt can include more model parameters without additional computational cost. Variational MaxEnt was also implemented to reweight the disease trajectory (See details in supporting information FIG. S2).  

\subsection*{Example System 3: MBP Fragment Molecular Dynamics}

Finally, we consider an application from biophysical modeling of the myelin basic protein (MBP) epitope fragment.
MBP is a common autoimmune target for the disease multiple scelrosis (MS)\cite{bielekova2000encephalitogenic}. Spyranti et al.\cite{spyranti2010nmr} characterized the specific region of MBP (83-99) that is the binding epitope for T-cell receptor recognition with solution Nuclear Magnetic Resonance (NMR).
NMR provides per-atom chemical shifts, which are population averages of a measurement of an atoms' local environment\cite{Cavanagh1995}. However, we must infer a specific structure to understand the molecular biology of MBP.
In this example we use molecular dynamics as a prior model, the chemical shifts as maximum entropy restraints, and compute a posterior of protein configurations. MaxEnt analysis has been applied frequently already in molecular dynamics, although not this exact approach with uncertainty\cite{Rangan2018RestraintvsReweight}.

Our prior model is an empirical distribution consisting of MBP fragment atomic positions as sampled from molecular dynamics. The specific fragment sequence was ENPVVHFFKNIVTPRTP and the molecular dynamics was initialized from an extended conformation. Simulations was performed in NVT ensemble in Gromacs 2020\cite{abraham2015gromacs,lindahl2001gromacs,pall2015tackling,berendsen1995gromacs,spoel2005gromacs,pronk2013gromacs, lindahl_2020_3923645} with CHARMM27 force field at a density of 25 mg/ml\cite{mackerell1998all, mackerell2004extending}. The simulation duration was 1.3$\mu$s with frames saved for this analysis every 500ps. To compute the chemical shift, $g_k$, we use a graph neural network that can compute chemical shift from atomic positions\cite{Yang2021}. We only biased backbone HN atoms, due to their higher accuracy\cite{Yang2021}. The first 6 HN atoms were biased (NPVVHF), excluding the N-terminus. The remaining HN atom chemical shifts were unbiased (KNIVTRT). $P_0(\epsilon)$ was chosen to be a Laplace distribution with scale parameter 0.05 -- allowing a small amount of systematic disagreement.

\begin{figure}
    \centering
    \includegraphics[width=0.6\textwidth]{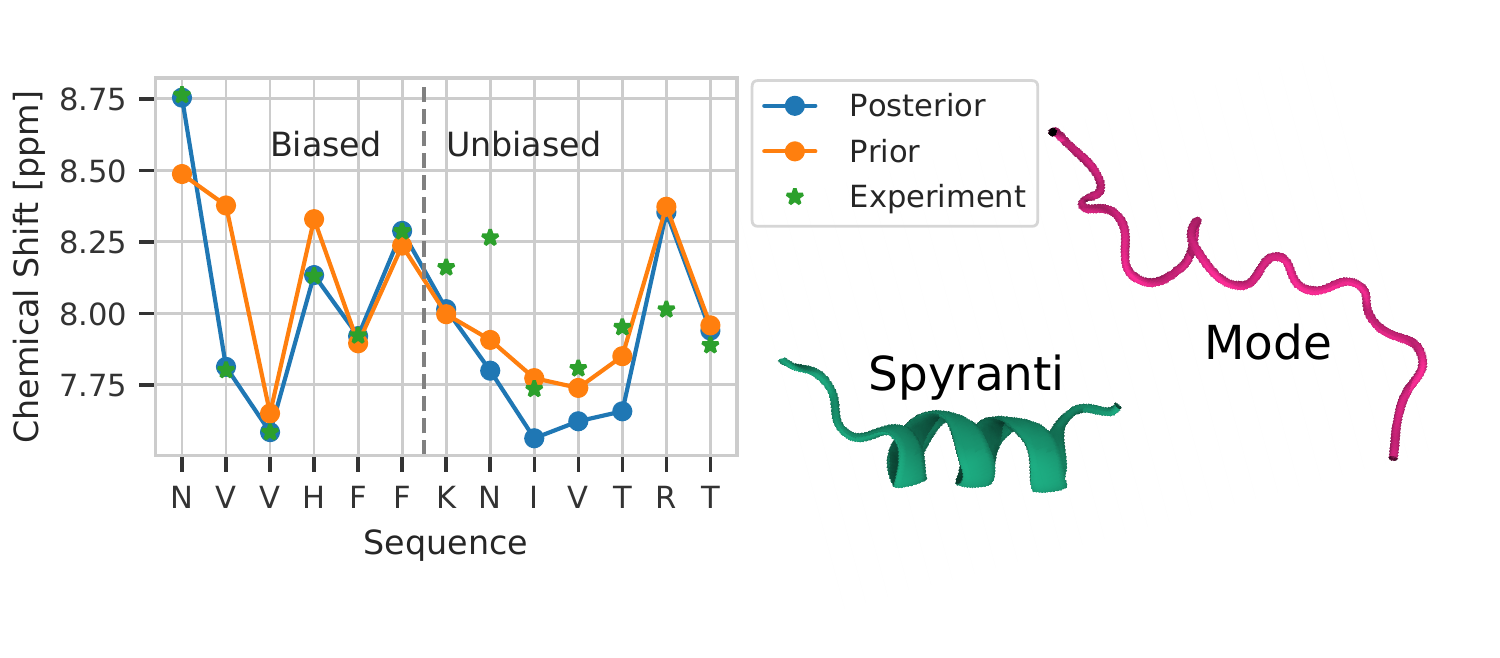}
    \caption{{\bf Maximum entropy reweighting of protein molecular dynamics simulation}. Molecular dynamics simulation of MBP epitope fragment was used to generate prior. Prior was reweighted with MaxEnt to have agreeing chemical shifts at indicated sequence positions. Chemical shifts are HN, as predicted by graph neural network\cite{Yang2021}. Experiments from Spyranti et al.\cite{spyranti2010nmr}. Mode is most weighted structure after MaxEnt. Spyranti is from NMR structure refinement.
    \label{fig:protein}}
\end{figure}

Figure \ref{fig:protein} shows the maximum entropy posterior average chemical shifts. As expected, exact agreement is found for the chemical shifts for which observations are provided. The posterior for which there are no observations follow the the prior closely (as expected in MaxEnt), although they move in the wrong direction at some positions. The protein structures are shown to the right of the plot. ``Spyranti'' is a representative structure from the deposited structure constructed by Spyranti et al.\cite{spyranti2010nmr} The posterior mode from MaxEnt is shown to the right. It has a helix, though not the $\alpha$-helix as shown in Spyranti. An advantage of this MaxEnt approach to analyzing NMR data is that there are 6500 structures in the posterior, whereas the traditional approach of NMR structure refinement results in 5-20 structures. This large distribution can then be used for other tasks with better calibration, such as finding drugs to target the protein structure, predicting protein-protein interfaces, and assessing structural properties.

\section*{Discussion}
We have presented MaxEnt reweighting as an inference method for altering an approximately-correct simulator to agree with observations. This method can be used on arbitrary simulators with arbitrary numbers of parameters, requiring only sufficient sampling of the prior distribution. The simulator need not have derivatives or tractable likelihoods. We demonstrated this by comparing with other SBI methods using three different simulators in different example contexts. The framework is particularly effective and robust when data is scarce or expensive (epidemic spreading being an archetypal example). MaxEnt provably changes the prior minimally to fit observations. While the method was initially developed for and particularly well-suited for molecular dynamics simulations---where experimental observations are much more costly and few in number compared to simulation---as demonstrated here, its applicability has potential for use in any setting of stochastic modeling where the derivative of the simulator's output with respect to latent variables is unavailable or intractable.

The approach to sampling described here is an implementation of variational inference to sample from the posterior. One could instead use Monte Carlo sampling. This would have the advantage of not requiring prior distribution derivatives, but since the derivatives here are closed-form it is computationally convenient to use importance sampling. MaxEnt's advantages over other widely used SBI methods, such as SNL, are that it is simple to implement, requires no hyperparameter choices like a neural network design, and can fit observations in expectation.

\section*{Methods}

\subsection*{Point Particle Gravitation Simulation}
For this simulation, the prior parameter distribution was taken as a multivariate normal distribution centered at $\{m_1=85,m_2=40,m_
3=70,v_{0x}=12,v_{0y}=-30\}$, with covariance matrix $\mathbf{I}\times 50$. This wide prior was chosen  because MaxEnt needs parameter support that overlaps with the observations we would like to fit. Fitting was done using the SBI package for Python~\cite{tejero-cantero2020sbi} with the SNL method,~\cite{papamakarios2019SNLE} and a custom implementation of MaxEnt reweighting using  Keras~\cite{tensorflow2015-whitepaper,chollet2015keras}. Both methods used 2048 prior samples for fitting. SNL used default parameters from the SBI package~\cite{tejero-cantero2020sbi} and MaxEnt used the Adam optimizer~\cite{KingmaAdam2014} with a learning rate of 0.0001 with mean squared error for 30000 epochs and batch size 2048.

\subsection*{Epidemiology Modeling}
 Epidemic spreading in networks can be modeled as a reaction-diffusion process. The reaction corresponds to an infection caused by interactions of subjects within a fully-mixed region or patch of varying granularities (a meta-population), while diffusion corresponds to movement of people (of various infection states) between patches~~\cite{gomez2018critical}. In this example, the meta-population system is comprised of three isolated local populations (patches) connected via flows corresponding to migrating individuals.  The spreading process is represented through a temporally discretized ODE that includes the spatial distribution of the population as well as their mobility patterns\cite{arenas2020mathematical}.

In our simulation, the infection begins in patch 1, propagating to the other two patches according to a synthetic mobility matrix. This mobility matrix was randomly generated with dominating diagonal elements to satisfy the fully-mixed region assumption. Five uniformly random data points within the first half of the trajectory of the compartment {\bf I} in patch 1 were considered as observations with a 5 percent random additive noise and Laplace prior of 0.01 (shown as restraints in Figure \ref{fig:Figure_4}a). The true parameters for the reference epidemic trajectory are: $\{start_{I}=0.02, start_{A}=0.05, E_{period}=7, A_{period}=5, I_{period}=14\}$. Parameters for this simulation were asymptomatic, infectious and exposed periods along with the fractional starting values for {\bf I} and {\bf A} compartments. The prior parameter distribution were taken as a truncated-normal  distribution centered at $\{start_{I}=0.001, start_{A}=0.001, E_{period}=2, A_{period}=2, I_{period}=10\}$, with variances of $\{0.8, 0.8, 1, 4, 5\}$, respectively. For the simulation, the pyABC~\cite{klinger2018pyabc} package was used with default parameters, and the same MaxEnt implementation was used with the Adam optimizer, a learning rate of 0.1, and loss of mean squared error for 1000 epochs with a batch size of 8192.

\subsection*{MBP Fragment Molecular Dynamics}
Molecular dynamics was done with Gromacs 2020.03 \cite{abraham2015gromacs,lindahl2001gromacs,pall2015tackling,berendsen1995gromacs,spoel2005gromacs,pronk2013gromacs, lindahl_2020_3923645} as driven by GromacsWrapper\cite{gromacswrapper} with a timestep of 2 fs. MBP initial structure were generated with PeptideBuilder\cite{peptidebuilder2013} and Packmol\cite{MARTINEZ2009}. The CHARMM27 forcefield was used for\cite{mackerell1998all, mackerell2004extending}. Canonical Sampling through Velocity Rescaling (CSVR) thermostat was used\cite{Bussi2007}. Long-range electrostatic forces were calculated with the particle mesh Ewald method\cite{essmann1995a}. Shifted Van der Waals and short-range electrostatics were used with a cutoff distance of 1 nm. Hydrogen containing covalent bonds were constrained using the LINear Constraint Solver (LINCS) algorithm\cite{Hess1997}. MaxEnt implementation as described above was used with 500 epochs in Adam optimizer with learning rate of 0.1.

\subsection*{Acknowledgements}
Funding for this research was provided by National Science Foundation under Grant No 2029095.

\subsection*{Code Availability Statement}

Code for this work is available at \url{https://github.com/ur-whitelab/maxent}. The SEAIR model implementation used in this work is publicly available as a python package at \url{https://github.com/ur-whitelab/py0}.

\bibliography{natcompsci}

\begin{thebibliography}{10}

\bibitem{Cranmer2020}
Kyle Cranmer, Johann Brehmer, and Gilles Louppe.
\newblock {The frontier of simulation-based inference}.
\newblock {\em Proceedings of the National Academy of Sciences}, page
  201912789, may 2020.

\bibitem{rubin1984bayesianly}
Donald~B Rubin.
\newblock Bayesianly justifiable and relevant frequency calculations for the
  applies statistician.
\newblock {\em The Annals of Statistics}, pages 1151--1172, 1984.

\bibitem{beaumont2002approximate}
Mark~A Beaumont, Wenyang Zhang, and David~J Balding.
\newblock Approximate bayesian computation in population genetics.
\newblock {\em Genetics}, 162(4):2025--2035, 2002.

\bibitem{diggle1984monte}
Peter~J Diggle and Richard~J Gratton.
\newblock Monte carlo methods of inference for implicit statistical models.
\newblock {\em Journal of the Royal Statistical Society: Series B
  (Methodological)}, 46(2):193--212, 1984.

\bibitem{BussiRNA}
Sabine Reißer, Silvia Zucchelli, Stefano Gustincich, and Giovanni Bussi.
\newblock {Conformational ensembles of an RNA hairpin using molecular dynamics
  and sparse NMR data}.
\newblock {\em Nucleic Acids Research}, 48(3):1164--1174, 12 2019.

\bibitem{Sormanni2017}
Pietro Sormanni, Damiano Piovesan, Gabriella~T Heller, Massimiliano Bonomi,
  Predrag Kukic, Carlo Camilloni, Monika Fuxreiter, Zsuzsanna Dosztanyi,
  Rohit~V Pappu, M~Madan Babu, Sonia Longhi, Peter Tompa, A~Keith Dunker,
  Vladimir~N Uversky, Silvio C~E Tosatto, and Michele Vendruscolo.
\newblock {Simultaneous quantification of protein order and disorder}.
\newblock {\em Nat. Chem. Biol.}, 13(4):339--342, 2017.

\bibitem{bonomi2017principles}
Massimiliano Bonomi, Gabriella~T Heller, Carlo Camilloni, and Michele
  Vendruscolo.
\newblock Principles of protein structural ensemble determination.
\newblock {\em Current opinion in structural biology}, 42:106--116, 2017.

\bibitem{Olsson2013}
Simon Olsson, Jes Frellsen, Wouter Boomsma, Kanti~V Mardia, and Thomas.
  Hamelryck.
\newblock {Inference of structure ensembles of flexible biomolecules from
  sparse, averaged data.}
\newblock {\em PLoS One}, 8(11):e79439, 2013.

\bibitem{Amirkulova2019}
Dilnoza~B Amirkulova and Andrew~D White.
\newblock Recent advances in maximum entropy biasing techniques for molecular
  dynamics.
\newblock {\em Molecular Simulation}, 45(14-15):1285--1294, 2019.

\bibitem{Pitera2012}
Jed~W Pitera and John~D Chodera.
\newblock {On the Use of Experimental Observations to Bias Simulated
  Ensembles.}
\newblock {\em Journal of Chemical Theory and Computation}, 8(10):3445--3451,
  2012.

\bibitem{berger1996maximum}
Adam Berger, Stephen~A Della~Pietra, and Vincent~J Della~Pietra.
\newblock A maximum entropy approach to natural language processing.
\newblock {\em Computational linguistics}, 22(1):39--71, 1996.

\bibitem{Roux2013statistical}
Beno{\^\i}t Roux and Jonathan Weare.
\newblock On the statistical equivalence of restrained-ensemble simulations
  with the maximum entropy method.
\newblock {\em The Journal of chemical physics}, 138(8):02B616, 2013.

\bibitem{de2018introduction}
Andrea De~Martino and Daniele De~Martino.
\newblock An introduction to the maximum entropy approach and its application
  to inference problems in biology.
\newblock {\em Heliyon}, 4(4):e00596, 2018.

\bibitem{banavar2010applications}
Jayanth~R Banavar, Amos Maritan, and Igor Volkov.
\newblock Applications of the principle of maximum entropy: from physics to
  ecology.
\newblock {\em Journal of Physics: Condensed Matter}, 22(6):063101, 2010.

\bibitem{wilson2020bayesian}
Andrew~Gordon Wilson and Pavel Izmailov.
\newblock Bayesian deep learning and a probabilistic perspective of
  generalization.
\newblock {\em arXiv preprint arXiv:2002.08791}, 2020.

\bibitem{Jaynes1957MaxEnt}
Edwin~T Jaynes.
\newblock Information theory and statistical mechanics.
\newblock {\em Physical review}, 106(4):620, 1957.

\bibitem{islam2013structural}
Shahidul~M Islam, Richard~A Stein, Hassane~S Mchaourab, and Beno{\^\i}t Roux.
\newblock Structural refinement from restrained-ensemble simulations based on
  epr/deer data: application to t4 lysozyme.
\newblock {\em The journal of physical chemistry B}, 117(17):4740--4754, 2013.

\bibitem{Vendruscolo2013molecular}
Andrea Cavalli, Carlo Camilloni, and Michele Vendruscolo.
\newblock Molecular dynamics simulations with replica-averaged structural
  restraints generate structural ensembles according to the maximum entropy
  principle.
\newblock {\em The Journal of chemical physics}, 138(9):03B603, 2013.

\bibitem{Boomsma2014combining}
Wouter Boomsma, Jesper Ferkinghoff-Borg, and Kresten Lindorff-Larsen.
\newblock Combining experiments and simulations using the maximum entropy
  principle.
\newblock {\em PLoS Comput Biol}, 10(2):e1003406, 2014.

\bibitem{white2014efficient}
Andrew~D White and Gregory~A Voth.
\newblock Efficient and minimal method to bias molecular simulations with
  experimental data.
\newblock {\em Journal of chemical theory and computation}, 10(8):3023--3030,
  2014.

\bibitem{beauchamp2014bayesian}
Kyle~A Beauchamp, Vijay~S Pande, and Rhiju Das.
\newblock Bayesian energy landscape tilting: towards concordant models of
  molecular ensembles.
\newblock {\em Biophysical journal}, 106(6):1381--1390, 2014.

\bibitem{rozycki2011saxs}
Bartosz R{\'o}{\.z}ycki, Young~C Kim, and Gerhard Hummer.
\newblock Saxs ensemble refinement of escrt-iii chmp3 conformational
  transitions.
\newblock {\em Structure}, 19(1):109--116, 2011.

\bibitem{leung2016rigorous}
Hoi Tik~Alvin Leung, Olivier Bignucolo, Regula Aregger, Sonja~A Dames, Adam
  Mazur, Simon Berneche, and Stephan Grzesiek.
\newblock A rigorous and efficient method to reweight very large conformational
  ensembles using average experimental data and to determine their relative
  information content.
\newblock {\em Journal of chemical theory and computation}, 12(1):383--394,
  2016.

\bibitem{choy2001calculation}
Wing-Yiu Choy and Julie~D Forman-Kay.
\newblock Calculation of ensembles of structures representing the unfolded
  state of an sh3 domain.
\newblock {\em Journal of molecular biology}, 308(5):1011--1032, 2001.

\bibitem{bernado2007structural}
Pau Bernad{\'o}, Efstratios Mylonas, Maxim~V Petoukhov, Martin Blackledge, and
  Dmitri~I Svergun.
\newblock Structural characterization of flexible proteins using small-angle
  x-ray scattering.
\newblock {\em Journal of the American Chemical Society}, 129(17):5656--5664,
  2007.

\bibitem{berlin2013recovering}
Konstantin Berlin, Carlos~A Castaneda, Dina Schneidman-Duhovny, Andrej Sali,
  Alfredo Nava-Tudela, and David Fushman.
\newblock Recovering a representative conformational ensemble from
  underdetermined macromolecular structural data.
\newblock {\em Journal of the American Chemical Society}, 135(44):16595--16609,
  2013.

\bibitem{bertini2010conformational}
Ivano Bertini, Andrea Giachetti, Claudio Luchinat, Giacomo Parigi, Maxim~V
  Petoukhov, Roberta Pierattelli, Enrico Ravera, and Dmitri~I Svergun.
\newblock Conformational space of flexible biological macromolecules from
  average data.
\newblock {\em Journal of the American Chemical Society}, 132(38):13553--13558,
  2010.

\bibitem{pelikan2009structure}
Martin Pelikan, Greg~L Hura, and Michal Hammel.
\newblock Structure and flexibility within proteins as identified through small
  angle x-ray scattering.
\newblock {\em General physiology and biophysics}, 28(2):174, 2009.

\bibitem{shaw2010atomic}
David~E Shaw, Paul Maragakis, Kresten Lindorff-Larsen, Stefano Piana, Ron~O
  Dror, Michael~P Eastwood, Joseph~A Bank, John~M Jumper, John~K Salmon, Yibing
  Shan, et~al.
\newblock Atomic-level characterization of the structural dynamics of proteins.
\newblock {\em Science}, 330(6002):341--346, 2010.

\bibitem{Bottaro2020BayesMaxEnt}
Sandro Bottaro, Tone Bengtsen, and Kresten Lindorff-Larsen.
\newblock {Integrating Molecular Simulation and Experimental Data: A
  Bayesian/Maximum Entropy Reweighting Approach}.
\newblock {\em Methods in Molecular Biology}, 2112:219--240, 2020.

\bibitem{Bradshaw2020DeuteriumMaxEnt}
Richard~T. Bradshaw, Fabrizio Marinelli, Jos{\'{e}}~D. Faraldo-G{\'{o}}mez, and
  Lucy~R. Forrest.
\newblock {Interpretation of HDX Data by Maximum-Entropy Reweighting of
  Simulated Structural Ensembles}.
\newblock {\em Biophysical Journal}, 118(7):1649--1664, apr 2020.

\bibitem{Lou2018HistogramMaxEnt}
Hongfeng Lou and Robert~I. Cukier.
\newblock {Reweighting ensemble probabilities with experimental histogram data
  constraints using a maximum entropy principle}.
\newblock {\em The Journal of Chemical Physics}, 149(23):234106, dec 2018.

\bibitem{Cesari2018MaxEntHowTo}
Andrea Cesari, Sabine Rei{\ss}er, and Giovanni Bussi.
\newblock {Using the Maximum Entropy Principle to Combine Simulations and
  Solution Experiments}.
\newblock {\em Computation}, 6(1):15, feb 2018.

\bibitem{Rangan2018RestraintvsReweight}
Ramya Rangan, Massimiliano Bonomi, Gabriella~T. Heller, Andrea Cesari, Giovanni
  Bussi, and Michele Vendruscolo.
\newblock {Determination of Structural Ensembles of Proteins: Restraining vs
  Reweighting}.
\newblock {\em Journal of Chemical Theory and Computation}, 14(12):6632--6641,
  dec 2018.

\bibitem{Blum2010SIRABC}
Michael G~B Blum and Chi Tran.
\newblock {HIV with contact tracing: a case study in approximate Bayesian
  computation}.
\newblock {\em Biostatistics}, 11(4):644--660, 2010.

\bibitem{Toni2009SIRABC}
Tina Toni, David Welch, Natalja Strelkowa, Andreas Ipsen, and Michael~P.H
  Stumpf.
\newblock {Approximate Bayesian computation scheme for parameter inference and
  model selection in dynamical systems}.
\newblock {\em Journal of The Royal Society Interface}, 6(31):187--202, 2009.

\bibitem{Kypraios2017SIRABC}
Theodore Kypraios, Peter Neal, and Dennis Prangle.
\newblock {A tutorial introduction to Bayesian inference for stochastic
  epidemic models using Approximate Bayesian Computation}.
\newblock {\em Mathematical Biosciences}, 287:42--53, 2017.

\bibitem{papamakarios2019SNLE}
George Papamakarios, David~C. Sterratt, and Iain Murray.
\newblock Sequential neural likelihood: Fast likelihood-free inference with
  autoregressive flows.
\newblock {\em ArXiv}, 2019.

\bibitem{Gordon2020BayesModelAverage}
Andrew Gordon and Wilson~Pavel Izmailov.
\newblock {Bayesian Deep Learning and a Probabilistic Perspective of
  Generalization}.
\newblock {\em Arxiv}, 2020.

\bibitem{cesari2016MaxEntUncertainty}
Andrea Cesari, Alejandro Gil-Ley, and Giovanni Bussi.
\newblock Combining simulations and solution experiments as a paradigm for rna
  force field refinement.
\newblock {\em Journal of chemical theory and computation}, 12(12):6192--6200,
  2016.

\bibitem{Tokdar2010ImportanceReview}
Surya~T. Tokdar and Robert~E. Kass.
\newblock {Importance sampling: a review}.
\newblock {\em Wiley Interdisciplinary Reviews: Computational Statistics},
  2(1):54--60, jan 2010.

\bibitem{hummer2015bayesian}
Gerhard Hummer and J{\"u}rgen K{\"o}finger.
\newblock Bayesian ensemble refinement by replica simulations and reweighting.
\newblock {\em The Journal of chemical physics}, 143(24):12B634\_1, 2015.

\bibitem{zhou2020pneumonia}
Peng Zhou, Xing-Lou Yang, Xian-Guang Wang, Ben Hu, Lei Zhang, Wei Zhang,
  Hao-Rui Si, Yan Zhu, Bei Li, Chao-Lin Huang, et~al.
\newblock A pneumonia outbreak associated with a new coronavirus of probable
  bat origin.
\newblock {\em nature}, 579(7798):270--273, 2020.

\bibitem{wu2020new}
Fan Wu, Su~Zhao, Bin Yu, Yan-Mei Chen, Wen Wang, Zhi-Gang Song, Yi~Hu, Zhao-Wu
  Tao, Jun-Hua Tian, Yuan-Yuan Pei, et~al.
\newblock A new coronavirus associated with human respiratory disease in china.
\newblock {\em Nature}, 579(7798):265--269, 2020.

\bibitem{Lipsitch2020COVIDLetter}
Marc Lipsitch, David~L. Swerdlow, and Lyn Finelli.
\newblock {Defining the Epidemiology of Covid-19 — Studies Needed}.
\newblock {\em New England Journal of Medicine}, 382(13):1194--1196, mar 2020.

\bibitem{bielekova2000encephalitogenic}
Bibiana Bielekova, Bonnie Goodwin, Nancy Richert, Irene Cortese, Takayuki
  Kondo, Ghazaleh Afshar, Bruno Gran, Joan Eaton, Jack Antel, Joseph~A Frank,
  et~al.
\newblock Encephalitogenic potential of the myelin basic protein peptide (amino
  acids 83--99) in multiple sclerosis: results of a phase ii clinical trial
  with an altered peptide ligand.
\newblock {\em Nature medicine}, 6(10):1167--1175, 2000.

\bibitem{spyranti2010nmr}
Zinovia Spyranti, Theodore Tselios, George Deraos, John Matsoukas, and
  Georgios~A Spyroulias.
\newblock Nmr structural elucidation of myelin basic protein epitope 83--99
  implicated in multiple sclerosis.
\newblock {\em Amino acids}, 38(3):929--936, 2010.

\bibitem{Cavanagh1995}
John Cavanagh, Wayne~J Fairbrother, Arthur~G Palmer~III, and Nicholas~J
  Skelton.
\newblock {\em Protein NMR spectroscopy: principles and practice}.
\newblock Elsevier, 1995.

\bibitem{abraham2015gromacs}
Mark~James {Abraham}, Teemu {Murtola}, Roland {Schulz}, Szilárd {Páll},
  Jeremy~C. {Smith}, Berk {Hess}, and Erik {Lindahl}.
\newblock Gromacs: High performance molecular simulations through multi-level
  parallelism from laptops to supercomputers.
\newblock {\em SoftwareX}, 1:19--25, 2015.

\bibitem{lindahl2001gromacs}
E.~{Lindahl}, B~{Hess}, and D.~van~der {Spoel}.
\newblock Gromacs 3.0: a package for molecular simulation and trajectory
  analysis.
\newblock {\em Journal of Molecular Modeling}, 7(8):306--317, 2001.

\bibitem{pall2015tackling}
Szilárd {Páll}, Mark~James {Abraham}, Carsten {Kutzner}, Berk {Hess}, and
  Erik {Lindahl}.
\newblock Tackling exascale software challenges in molecular dynamics
  simulations with gromacs.
\newblock {\em arXiv preprint arXiv:1506.00716}, pages 3--27, 2015.

\bibitem{berendsen1995gromacs}
H.J.C. {Berendsen}, D.~van~der {Spoel}, and R.~van {Drunen}.
\newblock Gromacs: A message-passing parallel molecular dynamics
  implementation.
\newblock {\em Computer Physics Communications}, 91(1):43--56, 1995.

\bibitem{spoel2005gromacs}
David Van~Der {Spoel}, Erik {Lindahl}, Berk {Hess}, Gerrit {Groenhof}, Alan~E.
  {Mark}, and Herman J.~C. {Berendsen}.
\newblock Gromacs: Fast, flexible, and free.
\newblock {\em Journal of Computational Chemistry}, 26(16):1701--1718, 2005.

\bibitem{pronk2013gromacs}
Sander {Pronk}, Szilárd {Páll}, Roland {Schulz}, Per {Larsson}, Pär
  {Bjelkmar}, Rossen {Apostolov}, Michael~R. {Shirts}, Jeremy~C. {Smith},
  Peter~M. {Kasson}, David van~der {Spoel}, Berk {Hess}, and Erik {Lindahl}.
\newblock Gromacs 4.5.
\newblock {\em Bioinformatics}, 29(7):845--854, 2013.

\bibitem{lindahl_2020_3923645}
Lindahl, Abraham, Hess, and van~der Spoel.
\newblock Gromacs 2020.3 source code, July 2020.

\bibitem{mackerell1998all}
Alex~D MacKerell~Jr, Donald Bashford, MLDR Bellott, Roland~Leslie Dunbrack~Jr,
  Jeffrey~D Evanseck, Martin~J Field, Stefan Fischer, Jiali Gao, H~Guo, Sookhee
  Ha, et~al.
\newblock All-atom empirical potential for molecular modeling and dynamics
  studies of proteins.
\newblock {\em The journal of physical chemistry B}, 102(18):3586--3616, 1998.

\bibitem{mackerell2004extending}
Alexander~D Mackerell~Jr, Michael Feig, and Charles~L Brooks~III.
\newblock Extending the treatment of backbone energetics in protein force
  fields: Limitations of gas-phase quantum mechanics in reproducing protein
  conformational distributions in molecular dynamics simulations.
\newblock {\em Journal of computational chemistry}, 25(11):1400--1415, 2004.

\bibitem{Yang2021}
Ziyue Yang, Maghesree Chakraborty, and Andrew~D. White.
\newblock Predicting chemical shifts with graph neural networks.
\newblock {\em Chem. Sci.}, 12:10802--10809, 2021.

\bibitem{tejero-cantero2020sbi}
Alvaro Tejero-Cantero, Jan Boelts, Michael Deistler, Jan-Matthis Lueckmann,
  Conor Durkan, Pedro~J. Gonçalves, David~S. Greenberg, and Jakob~H. Macke.
\newblock sbi: A toolkit for simulation-based inference.
\newblock {\em Journal of Open Source Software}, 5(52):2505, 2020.

\bibitem{tensorflow2015-whitepaper}
Mart\'{\i}n Abadi, Ashish Agarwal, Paul Barham, Eugene Brevdo, Zhifeng Chen,
  Craig Citro, Greg~S. Corrado, Andy Davis, Jeffrey Dean, Matthieu Devin,
  Sanjay Ghemawat, Ian Goodfellow, Andrew Harp, Geoffrey Irving, Michael Isard,
  Yangqing Jia, Rafal Jozefowicz, Lukasz Kaiser, Manjunath Kudlur, Josh
  Levenberg, Dandelion Man\'{e}, Rajat Monga, Sherry Moore, Derek Murray, Chris
  Olah, Mike Schuster, Jonathon Shlens, Benoit Steiner, Ilya Sutskever, Kunal
  Talwar, Paul Tucker, Vincent Vanhoucke, Vijay Vasudevan, Fernanda Vi\'{e}gas,
  Oriol Vinyals, Pete Warden, Martin Wattenberg, Martin Wicke, Yuan Yu, and
  Xiaoqiang Zheng.
\newblock {TensorFlow}: Large-scale machine learning on heterogeneous systems,
  2015.
\newblock Software available from tensorflow.org.

\bibitem{chollet2015keras}
François Chollet.
\newblock keras.
\newblock \url{https://github.com/fchollet/keras}, 2015.

\bibitem{KingmaAdam2014}
Diederik~P. Kingma and Jimmy Ba.
\newblock Adam: A method for stochastic optimization.
\newblock {\em CoRR}, abs/1412.6980, 2014.

\bibitem{gomez2018critical}
Jes{\'u}s G{\'o}mez-Gardenes, David Soriano-Panos, and Alex Arenas.
\newblock Critical regimes driven by recurrent mobility patterns of
  reaction--diffusion processes in networks.
\newblock {\em Nature Physics}, 14(4):391--395, 2018.

\bibitem{arenas2020mathematical}
Alex Arenas, Wesley Cota, Jesus Gomez-Gardenes, Sergio G{\'o}mez, Clara
  Granell, Joan~T Matamalas, David Soriano-Panos, and Benjamin Steinegger.
\newblock A mathematical model for the spatiotemporal epidemic spreading of
  covid19.
\newblock {\em MedRxiv}, 2020.

\bibitem{klinger2018pyabc}
Emmanuel Klinger, Dennis Rickert, and Jan Hasenauer.
\newblock {pyABC: distributed, likelihood-free inference}.
\newblock {\em Bioinformatics}, 34(20):3591--3593, 05 2018.

\bibitem{gromacswrapper}
Oliver Beckstein.
\newblock Gromacswrapper, 2017.

\bibitem{peptidebuilder2013}
Matthew~Z. Tien, Dariya~K. Sydykova, Austin~G. Meyer, and Claus~O. Wilke.
\newblock Peptidebuilder: A simple python library to generate model peptides,
  2013.

\bibitem{MARTINEZ2009}
L~Mart{\'{i}}nez, R~Andrade, E.~G. Birgin, and J.~M Mart{\'{i}}nez.
\newblock {Packmol: A Package for Building Initial Configurations forMolecular
  Dynamics Simulations}.
\newblock {\em Journal of Computational Chemistry}, 30(16):2157--2164, 2009.

\bibitem{Bussi2007}
Giovanni Bussi, Davide Donadio, and Michele Parrinello.
\newblock {Canonical sampling through velocity rescaling}.
\newblock {\em Journal of Chemical Physics}, 126(1):014101, 2007.

\bibitem{essmann1995a}
Ulrich {Essmann}, Lalith {Perera}, Max~L. {Berkowitz}, Tom {Darden}, Hsing
  {Lee}, and Lee~G. {Pedersen}.
\newblock A smooth particle mesh ewald method.
\newblock {\em Journal of Chemical Physics}, 103(19):8577--8593, 1995.

\bibitem{Hess1997}
Berk Hess, Henk Bekker, Herman J.~C. Berendsen, and Johannes G. E.~M. Fraaije.
\newblock {LINCS: A linear constraint solver for molecular simulations}.
\newblock {\em Journal of Computational Chemistry}, 18(12):1463--1472, 1997.

\end{thebibliography}
\bibliographystyle{unsrt}
\end{document}